\magnification=\magstep1  
\font\huge=cmr10 scaled \magstep2
\def\la{\lambda}   \overfullrule=0pt
\input amssym.def
\def\Z{{\Bbb Z}}  \def\Q{{\Bbb Q}}  
  
 \def\la{\lambda}  \def\ga{\gamma} 
       \def\i{{\rm i}} 
\def\si{\sigma} \def\eps{\epsilon}    
    \def\k{\kappa}     
  \def\L{{\Lambda}}

\font\smcap=cmcsc10


\centerline{{\bf \huge Algorithms for affine Kac-Moody algebras}}\bigskip
\bigskip

\centerline{{\smcap Terry Gannon}}\bigskip

\centerline{{\it Department of Mathematical Sciences, University of Alberta}}

\centerline{{\it Edmonton, Canada, T6G 2G1}}

\medskip\centerline{{e-mail: tgannon@math.ualberta.ca}}\bigskip\bigskip

\noindent{{\bf Abstract.}} Weyl groups are ubiquitous, and
efficient algorithms for them --- especially for the exceptional algebras --- are
clearly desirable. In this letter we provide several of these, addressing
practical concerns arising naturally for instance in computational
aspects of the study of affine algebras or Wess-Zumino-Witten (WZW)
conformal field theories.
We also discuss the efficiency and numerical accuracy of these
algorithms.



\bigskip\bigskip\noindent{{\bf 1. Introduction}}\medskip

\noindent This letter contains several formulas and algorithms involving the
(affine) Weyl groups, which play central roles in the
author's present work. Though some are surely known to other experts,
some are new, and they are all collected (and analysed) here to benefit also
the researchers and students less experienced with these matters.  An indication
of the need for good algorithms is
provided by the number of papers (see e.g.\ [1--4]) on it.

One place the finite Weyl groups arise is in the modular
transformation [5]
$$\chi_\la(-1/\tau)=\sum_{\mu\in P_+^k}S_{\la\mu}\,\chi_\mu(\tau)\eqno(1a)$$
 of affine characters,
where the sum is over the level $k$ integrable highest weights $\mu\in P_+^k$ for the
given affine Kac-Moody algebra $X_r^{(1)}$, for the $k$ satisfying $\la\in P_+^k$.
The level $k$ is a nonnegative integer, and the finite set $P_+^k$ is defined
in terms of the {\it colabels} $a_i^\vee$ by
$$P_+^k=\{(\la_1,\ldots,\la_r)\,|\,a_1^\vee\la_1+\cdots+a_r^\vee\la_r\le k,\
\la_i\in\Z_{\ge 0}\}\eqno(1b)$$
We identify the $r$-tuple with the combination $\la_1\L_1+\cdots+\la_r
\L_r$ in terms of the fundamental weights $\L_i$.
An expression for this unitary and symmetric {matrix} $S_{\la\mu}$ is
[5]
$$S_{\la\mu}={\k}^{-r/2}s\sum_{w\in {W}}({\rm det}\,w)\,\exp[-2\pi\i\,{
w(\la+\rho)\cdot(\mu+\rho)\over \k}]\eqno(1c)$$
where $\rho=(1,\ldots,1)$ and $\k=k+h^\vee$. The colabels $a_i^\vee$,
the {finite Weyl group} ${W}$, the number $s=\i^{\|\Delta_+\|}/\sqrt{\|Q^{\vee
*}/Q^\vee\|}$, the dot product,
and the dual Coxeter
number $h^\vee$, will be explicitly given below for each algebra.

The challenge raised by several researchers is to find an effective way of
computing this matrix $S$. For instance, in modular invariant classifications
[6], literally millions of $S$ entries typically must be computed
for each algebra.
The difficulty is that the Weyl group ${W}$
--- though finite --- can be very large and unwieldy. This is particularly
acute for the exceptional $X_r$: the Weyl groups of $G_2,F_4,E_6,E_7,E_8$
respectively have 12, $1\,152$, $51\,840$, $2\,903\,040$, and $696\,729\,600$
elements. 
We will show that this sum (1c) is easy to evaluate for the classical
$X_r^{(1)}$, and so a way to evaluate it for the exceptional ones would
be to find a large classical Weyl subgroup. Our algorithm involves computing
$O$((log$|{W}|)^2)$ trignometric functions and performing
$O$((log$|{W}|)^3)$ arithmetic operations. 

One of the many places in which
the matrix $S$ appears is {\it Verlinde's formula} [7] for the {\it
fusion coefficients} $N_{\la\mu}^\nu$. Write 0 for the weight
$(0,0,\ldots,0)\in P_+^k$. Then
$$N_{\la\mu}^\nu:=\sum_{\gamma\in P_+^k}\,{S_{\la\gamma}\,S_{\mu\gamma}\,S_{\nu
\gamma}^*\over S_{0\gamma}}\eqno(2a)$$
Verlinde's formula is not of much direct computational value:
to leading power in $k$, it has
$$\|P_+^k\|\approx {k^r\over r!\,a_1^\vee\cdots a_r^\vee}\eqno(2b)$$ 
terms. This also gives the size of  $S$. More effective is the
Kac-Walton formula [8,9], which expresses 
the fusion coefficients in terms of $X_r$ tensor product coefficients $T_{\la\mu}^\nu
={\rm mult}_\nu(L_\la\otimes L_\mu)$: 
$$N_{\la\mu}^\nu=\sum_{\hat{w}\in \widehat{W}}({\rm det}\,\hat{w})\,T_{\la\mu}^{\hat{w}.\nu}\eqno(2c)$$
where $\widehat{W}$ is the affine Weyl group; its action on
the weights $P_+^k$ will be discussed in \S2.2. 

Background material on simple Lie and affine algebras can be found in [8,10--12].
For an account of the rich role of the  matrix $S$ and fusion rules in rational
conformal field theory and elsewhere, see e.g.\ [13,14] and references therein.

In this paper we focus on three tasks: effective formulas for
computing $S$ entries; fast algorithms for finding the preferred affine
Weyl orbit representative $[\la]=\widehat{w}.\la\in P_+^k$; and super-fast
formulas for obtaining the parity $\epsilon(\la)={\rm det}(\widehat{w})$ for that representative, when
the weight $\widehat{w}.\la$ itself is not needed. In Section 4 we
sketch some applications.

\bigskip\noindent{\bf 2. Review and statement of the problems}\medskip

\noindent
Introduce the notation $P_\Z$ for all $\sum_i \la_i\L_i
=(\la_1,\ldots,\la_r)\in\Z^r$. It inherits
 the usual dot product of $P_+^k$. When the level $k$ is understood,
we'll write
$$\la_0=k-\sum_ia_i^\vee\la_i$$
Also, we'll abbreviate `algebra $X_r^{(1)}$ at level $k$' by `$X_{r,k}$'.
We will write `$\lfloor y\rfloor$' for the 
largest integer not greater than $y$ --- e.g.\ $\lfloor
\pi\rfloor=3=\lfloor 3\rfloor$. As always, $\k=k+h^\vee$.

\medskip \noindent{{\it 2.1. Nice weights and useful symmetries.}} If either
$\la$ or $\mu$ is a `nice' weight, then significant simplifications to (1c)
can be made. In this letter we are interested rather in {\it generic} weights
$\la,\mu$ in (1c) and so we will not devote much space reviewing these.

The denominator identity of Lie algebras permits us to express the sum (1c),
whenever $\la$ is a multiple of $\rho$, as a product over the positive roots:
$$S_{m\rho,\mu}=\k^{-r/2}|s|\prod_{\alpha\in\Delta_+}2\,\sin(\pi\,{(m+1)\,(\mu+\rho)\cdot
\alpha\over \k})\eqno(3)$$
The  most important choice is $m=0$, which yields a
product formula for the {\it quantum-dimensions} $S_{\la 0}/S_{00}$. 
For a rank $r$ Lie algebra
$X_r$, the Weyl group has approximate size $r!\approx \sqrt{2\pi r}(r/e)^r$,
while the number
of positive roots grows like $r^2$, so (3) is a significant simplification.

Remarkably, we will find that all entries of $S$ are 
about as accessible as $S_{\la 0}$, in particular, via equations with complexity
growing like $r^3$.

An important observation from (1c) is that the ratio 
$${S_{\la\mu}\over S_{0\mu}}={\rm ch}_\la(\exp[-2\pi\i\,{\mu+\rho\over \k}])
\eqno(4a)$$
is a finite Lie group character evaluated at an element of finite order [3].
Effective expressions exist for some ch$_\la$, implying that the corresponding
entries $S_{\la\mu}$ can be computed quickly. 
The weight multiplicities mult$_{L(\la)}(\beta)$ for simple
algebras of small rank and small weights are given explicitly in [15].
When $L(\la)$ has few dominant weights $\beta$, and each of these has a simple
Weyl orbit, then (4a) can be evaluated explicitly. For instance, for
$X_r=A_r$, when $\la$ is a fundamental weight $\Lambda_m$, we get
$${S_{\Lambda_m,\mu}\over S_{0\mu}}=\exp[2\pi\i\,m\,t(\mu+\rho)/(r+1)\k]
\sum_{1\le i_1<\cdots<i_m\le r+1}x_{i_1}\cdots x_{i_m}\eqno(4b)$$
where $t(\nu)=\sum_{i=1}^ri\nu_i$ and $x_i=\exp[-2\pi\i\,\mu^+[i]/\k]$
for orthogonal coordinates $\nu^+[i]=\sum_{j=i}^r(\nu_j+1)$.

Another application of (4a) is for the use of branching rules in
replacing $S_{\la\mu}/S_{0\mu}$ for $X_r^{(1)}$ with a sum (over
$\la'$) of $S'_{\la'\mu'}/S'_{0'\mu'}$ for a simpler algebra
$X'_s{}^{(1)}$ of equal rank. E.g.\ the embedding $D_8\subset E_8$ has
the branching rule $(\L_2')+(\L_8')=(\L_1)$ (see e.g.\ [30]), and
hence the ratio $S_{\L_1\mu}/S_{0\mu}$ equals the sum
$S'_{\L_2'\mu'}/S'_{0'\mu'}+S'_{\L_8'\mu'}/S'_{0'\mu'}$ for an
appropriate choice of $\mu'$ (depending only on $\mu$). Later we will work
this out explicitly, for both $D_4\subset F_4$ and $A_2\subset G_2$.

Related to this, for fixed $\ga\in P_+$, the ratios (4a) form a one-dimensional representation
of the fusion ring. That is,
$${S_{\la\ga}\over S_{0\ga}}\,{S_{\mu\ga}\over S_{0\ga}}=\sum_{\nu\in P_+}
N_{\la\mu}^\nu\,{S_{\nu\ga}\over S_{0\ga}}\eqno(5)$$
Using (2c), we can obtain $S$ entries for `larger' weights from those of smaller weights. 

Considerable simplifications [3] to the calculation of
ch$_\la(\exp[-2\pi\i\,{\mu\over \k}])$ occur when some Dynkin
labels $\mu_i$ of $\mu$ are 0. While this is not of direct value to us
(because of the `$+\rho$' in (4a)), it can be very useful as an
approximation when the level $k$ is large (see e.g.\ [6]).

Incidentally, the matrix $S$ obeys several symmetries. Best known are those
related to the symmetries of the extended Dynkin diagrams (see e.g.\ [12]). Thus
determining one $S$ entry automatically yields several others. For example, consider
$X_r=A_r$. Define an involution $C$ by $C(\la_1,\ldots,\la_r)=(\la_r,\ldots,
\la_2,\la_1)$, and an order-$(r+1)$ map $J$ by $J(\la_1,\ldots,\la_r)=
(\la_0,\la_1,\ldots,\la_{r-1})$ where $\la_0=k-\sum_{i=1}^r\la_i$. Then
both $C$ and $J$ are permutations on $P_+^k$, corresponding to the symmetries
of the extended Dynkin diagram of $A_r$. The various powers $J^i$ are
the {\it simple-currents} of $A_r^{(1)}$. We have $S_{C\la,\mu}=S_{\la,\mu}^*$ and
$$S_{J\la,\mu}=\exp[2\pi\i\,t(\mu)/(r+1)]\,S_{\la\mu}\eqno(6)$$
Similar equations hold for the other algebras (see e.g.\ [12]).

Another symmetry --- more complicated but more powerful --- is the Galois
symmetry [16] discussed in \S4.4 below. Making it more accessible is a
big motivation for the paper.

For the classical algebras, another symmetry ({\it rank-level duality} [17])
of the $S$ matrix tells us that small level acts like small rank. In particular,
the $S$ matrix for $\widehat{{\rm sl}(n)}_k$ and $\widehat{{\rm sl}(k)}_n$
are closely related, as are $\widehat{{\rm so}(n)}_k$ and
$\widehat{{\rm so}(k)}_n$, and $C_{r,k}$ and $C_{k,r}$. 

Finally, simplifications (called
{\it fixed-point factorisation} [18]) occur whenever $\la$
or $\mu$ is a fixed-point of a (nontrivial) simple-current --- the simple-currents
$J^i$ for $A_r^{(1)}$ are given above. If $\varphi$ is fixed by $J^d$,
then for any weight of $\la$ of $\widehat{{\rm sl}}(n)_k$ we can write
$S_{\varphi \la}$ as a product of $n/d$ $S$-entries for $\widehat{{\rm
sl}}(d)_{kd/n}$. For a very simple example, take
$\widehat{{\rm sl}}(n)_k$ when $n$ divides $k$: the unique $J$-fixed-point
is $\varphi=({k\over n},\ldots,{k\over n})$, and
$S_{\la,\varphi}=\pm({n\over k})^{(n-1)/2}$ or 0, depending on $\la$.

Incidentally, an analogue of fixed-point factorisation works for the
symmetries (called {\it conjugations}) of the unextended Dynkin
diagrams. For example, consider the $\Z_3$-symmetry {\it triality} of
$D_4$: a weight $\mu$ invariant under it obeys $\mu_1=\mu_3=\mu_4$,
and the $D_4^{(1)}$ ratio $S_{\la\mu}/S_{0\mu}$ for such a $\mu$
equals a sum $\sum_\la'S'_{\la'\mu'}/S_{0'\mu'}$ of $G_2^{(1)}$ ratios, where
$\oplus (\la')=(\la)$ are $G_2\subset D_4$ branching rules (see e.g.\
[30]), and $\mu'=(\mu_2,3\mu_1+2)\in P_+^{k+2}(G_2)$. Applications of
these fixed-point factorisations is given in [31].

In  practice the remarks of this subsection are quite effective. For example,
consider $A_{3,6}$. There are $7056$ $S$ entries: any involving
$J^j0$, $J^j(1,1,1)$, $J^j(2,2,2)$, $J^j\L_i$, $J^j(1,0,1)$, $J^j(1,1,0)$,
$C^iJ^j(2,0,0)$, or any $J^2$-fixed point, are immediately calculated from 
the above. Finally, exploiting the Galois
symmetry (see \S4.4 below), we reduce the calculation of the
$7056$
entries $S_{\la\mu}$ to precisely 1 less pleasant calculation: that of
$S_{(3,0,0),(3,0,0)}$. Of course, for large $k$ and $r$, most
weights won't be `nice' in this sense, and another approach is required.

\bigskip\noindent{{\it 2.2. Calculating Weyl orbits.}}
The affine Weyl group $\widehat{W}$ is generated by the reflections $r_i$ through simple
roots: explicitly, for any $\la\in P_\Z$, the Dynkin labels of
$r_i\la$ are
$$(r_i\la)_j=\la_j-\la_i A_{ji}\eqno(7a)$$
where $A_{ji}$ are entries of the Cartan matrix of $X_r^{(1)}$.
It is important that $\widehat{W}$ can be expressed as  the semidirect product of the finite Weyl group
with the coroot lattice interpreted as an additive group --- i.e.\
each $\hat{w}$ can be uniquely identified with a pair $(w,\alpha)$,
for $w\in W$ and $\alpha\in Q^\vee$, and $\hat{w}\la=w(\la)+\k
\alpha$. It is often
more convenient to express this action using the notation
$$\hat{w}.\la:= \hat{w}(\la+\rho)-\rho\eqno(7b)$$
Given any level $k$
weight $\la=(\la_1,\ldots,\la_r)\in P_\Z$, there are two possibilities: either
$\la+\rho$ is fixed by some 
 affine Weyl reflection $\hat{w}$, i.e.\ $\hat{w}.\la=\la$, 
 and we shall call $\la$ {\it null}; or the orbit $\widehat{W}.\la$ intersects the fundamental alcove
$P_+^k$ in precisely one point:
$$[\la]:=\hat{w}.\la\in P_+^k\eqno(7c)$$
Define the {\it parity} $\epsilon(\la)$ to be $0$ or ${\rm det}({w})$,
  depending on whether or not $\la$ is null.
This preferred orbit representative
$[\la]$, and/or the parity $\epsilon(\la)$, are often desired --- see 
\S4.

A very useful fact is that $\la$ will be null if $\la_i=-1$
for any $i=0,1,\ldots,r$.

A method for finding $[\la]$ and $\epsilon(\la)$ is proposed in [12]. Put $\epsilon=+1$ and $\mu=\la$.

\smallskip\item{(1)} If $\mu_i\ge 0$ for all $i=0,1,\ldots,r$, then $[\la]=\mu$
and $\epsilon(\la)=\epsilon$.

\item{(2)} Otherwise, let $0\le i\le r$ be the smallest index for which
$\mu_i<0$. Replace $\mu$ with $r_i.\mu$, and $\epsilon$ with $-\epsilon$.
Goto (1). \smallskip

It does not appear to
be known yet whether their method will always terminate, although in practice
it seems to. We give an alternative next section.

\bigskip\bigskip\noindent{{\bf 3. The formulas and algorithms}}\bigskip 

\noindent
The essence of the following $S$ matrix formulas is the observation that an alternating sum over the symmetric
group also occurs in determinants. The point is that determinants can be evaluated
very effectively, using Gaussian elimination, and is easy to implement on a
computer. By comparison, determinants are much more accessible than 
permanents.  
This suggests that we look no further than
(1c) for an effective algorithm.

The classical $S$ matrices all reduce to evaluating one or two determinants.
Our strategy for the exceptional algebras is to find a classical Weyl group
of small index. For example $W(D_8)$ is contained in $W(E_8)$ with index 135.
Once we find coset representatives for the 135 cosets in $W(D_8)\backslash
W(E_8)$, then the $E_8$ $S$ matrix would be the alternating sum of 135 $D_8$
$S$ matrix entries. Equally important, these coset representatives also permit effective
algorithms to find $[\la]$ and $\epsilon(\la)$, as we will see.

A priori,  it could be hard to find these coset representatives,
but here the task is made elementary by the following result [19] (there's
a typo in their definition of $D_\Psi$): 

\medskip\noindent{{\bf Theorem.}} {\it Let $\Psi$ be a subsystem of a root system
$\Phi$. Choose simple roots $\alpha\in J$ for $\Psi$ which are positive roots
$\Phi^+$ in $\Phi$ (this is always possible). Define $D_\Psi=\{w\in W(\Phi)\,|\,w(J)\subset
\Phi^+\}$. Then every element of $W(\Phi)$ can be uniquely expressed in the
form $d\,w'$ where $d\in D_\Psi$ and $w'\in W(\Psi)$. Furthermore, $d$ is the
unique element of minimal length in the coset $d\,W(\Psi)$.}\medskip

In particular, the set of all $d^{-1}$, for $d\in D_\Psi$, constitute the desired
coset representatives for $W(\Psi)\backslash W(\Phi)$. We chose the largest
classical subsystem possible for each of the exceptional root systems. The
results are explained in the following subsections.

To give the extreme example, we read off from Table 2 that the $W(D_8)\backslash W(E_8)$ coset
representative we call $c_{135}$ is the following composition of simple
reflections: 
$$r_7r_6r_5r_4r_3r_2r_1r_8r_5r_4r_3r_2r_6r_5r_4r_3r_8r_5r_4r_6
r_5r_8\ .$$
 It acts on a weight $\la\in P(E_8)$ as follows: write $\nu=c_{135}
.\la$, then using notation defined in \S3.E8 we get
$$\left(\matrix{\nu^+[1]\cr\vdots\cr\nu^+[8]\cr}\right)={1\over 4}\left(\matrix{
-1&1&1&1&1&1&-3&-1\cr
-1&1&1&1&1&-3&1&-1\cr
-1&1&1&1&-3&1&1&-1 \cr
-1&1&1&-3&1&1&1&-1 \cr
-1&1&-3&1&1&1&1&-1 \cr
-1&-3&1&1&1&1&1&-1 \cr
3&1&1&1&1&1&1&-1 \cr
-1&1&1&1&1&1&1&3 \cr
}\right)\left(\matrix{\la^+[1]\cr\vdots\cr\la^+[8]\cr}\right)$$

Root system data is explicitly given in e.g.\ [10,12]. For convenience
we reproduce it in the following subsections.

The best way to determine the parity $\epsilon(\la)$ of \S2.2 seems to be to use  (3):
$$\epsilon(\la)=\prod_{\alpha>0}{\rm sign}\{\sin(\pi\,{(\la+\rho)\cdot\alpha
\over \k})\}\eqno(8)$$
The (easy) proof that (8) holds is that the product of sine's can be expressed as
alternating sums over the Weyl group, thanks to equation (3),
and that when evaluated at any weight $\la\in P_+^k$ they yield a positive value
(proportional to the quantum-dimension). Equation (8) has been kicking
around for years --- see e.g.\ the authors earlier work, and also
[20]. Later in this section we make significant improvements to (8)
for all algebras except $A_r^{(1)}$.


We often require ordering lists of numbers
(lists whose length $N$ is approximately the rank $r$ of the algebra). Straightforward
ordering algorithms have a computing time growing like $N^2$, but more
sophisticated algorithms have order $N\,{\rm log}\,N$, and so are relevant
for large rank. See  [21] for details.

Many of our formulas involve taking determinants of $N\times N$ matrices,
where $N$ approximately equals the rank. Using Gaussian elimination,
this involves approximately $N^3$ operations. We will briefly discuss the error
analysis in \S4.1 --- see also [22].

Some of the following (namely the $S$ matrix for the classical algebras)
has appeared in [17], and B.\ Schellekens has used but not
published similar $S$ matrix formulas (see his webpage at
http://www.nikhef.nl/\~{}t58/kac.html).  Determinant formulas for the
characters of the classical Lie algebras go back to Weyl and are
discussed in \S24.2 of [23]. Other approaches to computing
generic $S$ entries can be found in [1,2,4].

\medskip\noindent{3.A. {\it The A-series.}}
The colabels $a_i^\vee$ for $X_r=A_r={\rm sl}(r+1)$ all equal $1$. The scale factor
in (1b) is $s=\i^{r(r+1)/2}/\sqrt{r+1}$, and $\k=k+r+1$.
To the weight $\la=(\la_1,\la_2,\ldots,\la_r)$ associate the orthogonal
coordinates $\la[\ell]=\sum_{i=\ell}^r\la_i$ for $\ell=1,\ldots,r+1$,
and write $t(\la)=\sum_{\ell=1}^r\ell\la_\ell$ and
$\la^+[j]:= (\la+\rho)[j]=r+1-j+\la[j]$. The finite Weyl group is
the symmetric group $S_{r+1}$, acting on $\la$ by permuting the components
$\la[\ell]$. The affine Weyl group also acts on a level $k$ weight $\la$ by
translation: $\la\mapsto\la+k\alpha $ for any $\alpha$ belonging to the
$A_r$ root lattice, i.e.\ $\alpha=(c_1,\ldots,c_{r+1})\in\Z^{r+1}$,
$\sum_ic_i=0$. Then
$$S_{\la\mu}=\k^{-r/2}s\exp[2\pi\i\,{t(\la+\rho)\,t(\mu+\rho)\over \k\,(r+1)}]\,
\,{\rm det}(\exp[-2\pi\i\,{\la^+[i]\,\mu^+[j]\over \k}])_{1\le i,j\le r+1}
\eqno(A.1)$$
`det' in (A.1) denotes the determinant of the $(r+1)\times(r+1)$ matrix
whose $(i,j)$-th entry is provided.

The fastest algorithm known to this author for finding the orbit representative
$[\la]$ of (7c) is as follows.\smallskip

\item{(i)} Write $x_\ell=\la^+[\ell]=r+1-\ell+\sum_{i=\ell}^{r}
\la_i$ for $\ell=1,2,\ldots,r$ and put $x_{r+1}=0$.

\item{(ii)} Reorder these so that $x_1>x_2>\cdots>x_r> x_{r+1}\ge 0$. 
If $x_1-x_{r+1}<\k$, then done; goto step (iv).

\item{(iii)} Otherwise for each $i=1,\ldots,(r+1)/2$ put
$m_i=\lfloor (x_i-x_{r+2-i})/\k\rfloor$; if $m_i$
is even let $x_i=x_i-m_i\k/2$ and $x_{r+2-i}=x_{r+2-i}+m_i\k/2$, while if $m_i$
is odd let  $x_i=x_{r+2-i}-(m_i+1)\k/2$ and $x_{r+2-i}=x_{i}+(m_i+1)\k/2$.
Return to step  (ii). 

\item{(iv)} The Dynkin labels $[\la]_i$ of the desired weight are
$$[\la]_i=y_{i}-y_{i+1}-1$$

The weight $\la$ is null iff this algorithm breaks down: i.e.\ two (or
more) 
$x_i$'s are equal. The parity
$\epsilon(\la)$ is the product of the signs of the permutations in
step (ii), together with $(-1)^m$'s from step (iii). An alternative is
provided in [24].

Equation (8) becomes
$$\epsilon(\la)={\rm sign}\prod_{1\le i<j\le r+1}\sin(\pi\,{\la^+[i]-\la^+[j]\over \k})
\eqno(A.2)$$

\medskip\noindent{3.B. {\it The B-series.}}
For $X_r=B_r={\rm so}(2r+1)$, $a_1^\vee=a_r^\vee=1$ and all other colabels equal $2$. The
scale factor is $s=
\i^{r^2}/2$, and $\k=k+2r-1$. To the weight $\la=(\la_1,\la_2,\ldots,\la_r)$ associate the orthogonal
coordinates $\la[\ell]=\sum_{i=\ell}^{r-1}\la_i+\la_r/2$ for $\ell=1,\ldots,r$
(put $\sum_{i=r}^{r-1}\la_i=0$)
and write $\la^+[j]:=(\la+\rho)[j]=r+{1\over 2}-j+\la[j]$.
The finite Weyl group is a group of order $2^r\cdot r!$,
acting on $\la$ by arbitrarily permuting and changing signs of the components
$\la[\ell]$. The affine Weyl group also acts on a level $k$ weight $\la$ by
translation: $\la\mapsto\la+k\alpha $ for any $\alpha$ belonging to the
$D_r$ root lattice, i.e.\ $\alpha=(c_1,\ldots,c_{r})\in\Z^{r}$ with
$\sum_ic_i$ is even. Then
$$S_{\la\mu}=\k^{-r/2}2^{r-1}\i^{r^2-r}\,{\rm det}(\sin(2\pi\,{\la^+[i]\,\mu^+[j]\over\k}))_{
1\le i,j\le r}\eqno(B.1)$$

The best algorithm known to this author for finding the orbit representative
$[\la]$ is:\smallskip

\item{(i)} Write $x_\ell=\la^+[\ell]=r-\ell+{1\over 2}+\sum_{i=\ell}^{r-1}
\la_i+\la_r/2$ for $\ell=1,2,\ldots,r$.

\item{(ii)} By adding the appropriate multiples of $\k$ to each
$x_\ell$, 
find the unique numbers $y_1,\ldots,y_{r}$ such that $x_\ell\equiv y_\ell$
(mod $\k$) and $-\k/2<y_\ell\le\k/2$. Let $m$ be the total number 
of $\k$ added: i.e.\ if $x_i=y_i+m_i\k$, then $m=\sum_{i=1}^rm_i$. We need
to know later whether $m$ is even or odd.

\item{(iii)} Replace each $y_i$ with its absolute value. 
Reorder these so that $y_1>y_2>\cdots>y_r> 0$.
 
\item{(iv)}  If $m$ is odd, replace $y_1$ with $\k-y_1$.

\item{(v)} Then the Dynkin labels $[\la]_i$ of the desired weight are
$$[\la]_i=y_{i}-y_{i+1}-1,\quad{\rm for}\ 1\le i<r,\qquad {\rm and}\ [\la]_r=2y_r-1$$
\smallskip

The weight $\la$ is null iff two of the $y_i$'s in (iii) are equal, or
if at least one $y_i$ equals 0. The parity $\epsilon(\la)$
is $(-1)^m$ times the sign of the product $\prod y_i$ in (ii), times the
sign of the permutation in (iii).

Equation (8) becomes (note the decoupling of the $i$th and $j$th terms here)
$$\epsilon(\la)={\rm sign}\prod_{1\le i\le r}\sin(\pi\,{\la^+[i]\over\k})\,
\prod_{1\le i<j\le r}\{\cos(2\pi\,{\la^+[i]\over \k})-\cos(2\pi\,{\la^+[j]
\over\k})\}\eqno(B.2)$$

where we used the identity $2\sin(x-y)\sin(x+y)=\cos(2y)-\cos(2x)$. 

\medskip\noindent{3.C. {\it The C-series.}}
The colabels for $X_r=C_r={\rm sp}(2r)$ all equal $1$. The scale
factor  is $s=\i^{r^2}/2^{r/2}$, and $\k=k+r+1$.
To the weight $\la=(\la_1,\la_2,\ldots,\la_r)$ associate the orthogonal
coordinates $\la[\ell]=\sum_{i=\ell}^r\la_i$ for $\ell=1,\ldots,r$, and write
$\la^+[\ell]=(\la+\rho)[\ell]=r+1-\ell+\la[\ell]$.
The finite Weyl group is a group of order $2^r\cdot r!$,
acting on $\la$ by arbitrarily permuting and changing signs of the components
$\la[\ell]$. The affine Weyl group also acts on a level $k$ weight $\la$ by
translation: $\la\mapsto\la+k\alpha $ for any $\alpha$ belonging to 
$(2\Z)\oplus\cdots\oplus(2\Z)$, i.e.\ $\alpha=(c_1,\ldots,c_{r})\in\Z^{r}$,
where each $c_i$ is even. Then
$$S_{\la\mu}=(2/\k)^{r/2}\i^{r^2-r}{\rm det}(\sin(\pi\,{\la^+[i]\,\mu^+[j]\over\k}))_{
1\le i,j\le r}\eqno(C.1)$$

The best algorithm known to this author for finding the orbit representative
$[\la]$ is:\smallskip

\item{(i)} Write $x_\ell=\la^+[\ell]=r+1-\ell+\sum_{i=\ell}^{r}
\la_i$.

\item{(ii)} By adding the appropriate multiples of $2\k$ to each $x_\ell$,
find the unique numbers $y_1,\ldots,y_{r}$ such that $x_\ell\equiv y_\ell$
(mod $2\k$) and $-\k<y_\ell\le\k$. 

\item{(iii)} Replace each $y_i$ with its absolute value. Reorder these so
that $\k\ge y_1>y_2>\cdots>y_r>0$. 

\item{(iv)} Then the Dynkin labels $[\la]_i$ of the desired weight are
(put $y_{r+1}=0$)
$$[\la]_i=y_{i}-y_{i+1}-1$$
\smallskip

The weight $\la$ is null iff some $y_i=0$, or two $y_i$'s are equal.
The parity $\epsilon(\la)$
is the sign of the product $\prod y_i$ in (ii), times the
sign of the permutation in (iii).

Equation (8) becomes (note the decoupling of the $i$th and $j$th terms here)
$$\eqalignno{\epsilon(\la)
=&\,{\rm sign}\prod_{1\le i\le r}\sin(\pi\,{\la^+[i]\over\k})\,
\prod_{1\le i<j\le r}\{\cos(\pi\,{\la^+[j]\over \k})-\cos(\pi\,{\la^+
[i]\over \k})\}&(C.2)}$$

\medskip\noindent{3.D. {\it The D-series.}}
For $X_r=D_r={\rm so}(2r)$, $a_1^\vee=a_{r-1}^\vee=a_r^\vee=1$ and all other colabels
are $2$. The scale factor in (1b) is $s=\i^{r(r-1)}/2$ and $\k=k+2r-2$.
To the weight $\la=(\la_1,\la_2,\ldots,\la_r)$ associate the orthogonal
coordinates $\la[\ell]=\sum_{i=\ell}^{r-1}\la_i+{\la_r-\la_{r-1}\over 2}$
for $\ell=1,\ldots,r$ ($\sum_{i=r}^{r-1}=0$), and define $\la^+[\ell]=(\la+\rho)[\ell]=r-\ell+\la[\ell]$.
The finite Weyl group has order $r!\,2^{r-1}$, acting
on $\la$ by permuting the components $\la[\ell]$ and changing an even number
of signs. The action of the affine Weyl group includes the translations
 $\la\mapsto\la+k\alpha $ for any $\alpha$ belonging to the
$D_r$ root lattice, i.e.\ $\alpha=(c_1,\ldots,c_{r})\in\Z^{r}$ where
$\sum_ic_i$ is even. Then
$$\eqalignno{S_{\la\mu}=\k^{-r/2}2^{r-2}\i^{r(r-1)}\,\{&{\rm det}
(\cos(2\pi\,{\la^+[i]\,\mu^+[j]\over \k}))_{1\le i,j\le r}&(D.1)\cr&+
(-\i)^r{\rm det}(\sin(2\pi\,{\la^+[i]\,\mu^+[j]\over\k}))_{
1\le i,j\le r}\}&\cr}$$

The best algorithm known to this author for finding the orbit representative
$[\la]$ is:\smallskip

\item{(i)} Write $x_\ell=\la^+[\ell]=r-\ell+\sum_{i=\ell}^{r-1}
\la_i+{\la_r-\la_{r-1}\over 2}$.

\item{(ii)} By adding the appropriate multiples of $\k$ to each $x_\ell$,
find the unique numbers $y_1,\ldots,y_{r}$ such that $x_\ell\equiv y_\ell$
(mod $\k$) and $-\k/2<y_\ell\le\k/2$.  Let $m$ be the total number 
of $\k$ added: i.e.\ if $x_i=y_i+m_i\k$, then $m=\sum_{i=1}^rm_i$. We need
to know later whether $m$ is even or odd.

\item{(iii)} Let $m'=\pm 1$ be the sign of $\prod_iy_i$ (take $m'=1$ if the
product is 0). Replace each $y_i$ with its absolute value $|y_i|$. Reorder
these so that $\k/2\ge y_1>y_2>\cdots>y_r\ge 0$. 

\item{(iv)} Replace $y_r$ with $m'y_r=\pm y_r$. If $m$ is odd, replace
$y_1$ with $\k-y_1$ and $y_r$ with $-y_r$.

\item{(v)} Then the Dynkin labels $[\la]_i$ of the desired weight are
$$[\la]_i=y_{i}-y_{i+1}-1\quad 1\le i\le r-1,\qquad[\la]_r=y_r+y_{r-1}-1$$
\smallskip

The weight is null iff two $y_i$'s in (iii) are equal. 
The parity $\epsilon(\la)$
is the sign of the permutation in (iii).

Equation (8) becomes (note the decoupling of the $i$th and $j$th terms here)
$$\eqalignno{\epsilon(\la)
=&\,{\rm sign}\prod_{1\le i<j\le
r}\{\cos(2\pi\,{\la^+[j]\over \k})-\cos(2\pi\,{\la^+[i]\over\k})\}&(D.2)}$$
Curiously, this sign has the interpretation as the sign of the
permutation which orders the numbers $\cos(2\pi\,\la^+[i]/\k)$ in increasing
order.

\medskip\noindent{3.E6. {\it The algebra $E_6$.}}
The colabels for $E_6$ are 1,2,3,2,1,2, respectively. The scale factor
in  (1b) is $s=1/\sqrt{3}$, and $\k=k+12$.
To the weight $\la=(\la_1,\la_2,\ldots,\la_6)$ associate the orthogonal
coordinates (interpret $\sum_{i=2}^1\la_i$ as 0)
$$\la[\ell]={\la_6-\la_2\over 2}+\sum_{i=2}^\ell\la_i\ {\rm for}\ \ell\le 5,
\quad\la[6]={5\la_2+3\la_6\over 2}+3\la_3+2\la_4+\la_5+2\la_1$$

The 27 coset representatives for $W(D_5)\backslash W(E_6)$ are given in Table 1,
recursively defined in terms of the reflections $r_i$, $i=1,\ldots,6$,
 through the simple roots $\alpha_i$. Then
$$\eqalignno{S_{\la\mu}={16\over\sqrt{3}\k^3}\sum_{\ell=1}^{27}({\rm det}\,c_\ell)\,
\{&\,{\rm det}
(\cos(2\pi\,{(c_\ell.\la)^+[i]\,\mu^+[j]\over \k}))_{1\le i,j\le 5}&(E6.1)\cr&
-\i\,{\rm det}(\sin(2\pi\,{(c_\ell.\la)^+[i]\,\mu^+[j]\over\k}))_{
1\le i,j\le 5}\}&\cr}$$
Recall that the numbers $(c_\ell.\la)^+[i]$ are the orthogonal
coordinates of the weight $c_\ell(\la+\rho)$.

The orbit representative $[\la]$ in (7c) can be found as follows.

\smallskip\item{(i)} Choose one of the 27 coset representatives $c_\ell$. Write
$z_i:=(c_\ell.\la)^+[6-i]$, $i=1,\ldots,5$, and $z_6=(c_\ell.\la)^+[6]$,
 for the orthogonal coordinates (reordered) of
$c_\ell(\la+\rho)$. (The `$[6-i]$' is chosen to make the first five
indices here consistent with those of $D_5$.)

\item{(ii)} Add an appropriate multiple $3n\k/2$ of $3\k/2$ to $z_6$
to get  $0\le x_6<3\k/2$; replace $x_1=n\k/2+z_1$, and
$x_i=-n\k/2+z_i$ for $2\le i\le 5$. Use the $D_5$ algorithm for the
present value of $\k$ to get
$|y_5|<y_4<y_3<y_2<y_1$ and $y_1+y_2<\k$.

\item{(iii)} Write $t=(-y_1-y_2-y_3-y_4+y_5+y_6)/2$. If $t>0$ and
$y_1+y_2+\cdots+y_6<2\k$, then the Dynkin labels of the desired weight
$[\la]$ are 
$$[\la]_1=t-1\ ,\ [\la]_6=y_4+y_5-1\ ,\ [\la]_i=y_{6-i}-y_{7-i}-1\qquad 2\le
i\le 5$$

\item{(iv)} Otherwise try the next $c_\ell$ in (i).\smallskip

Equation (8) becomes
$$\eqalignno{\epsilon(\la)={\rm sign} &\,\prod_{1\le i<j\le 5}\{\cos(2\pi{\la^+[i]\over
\k})-\cos(2\pi{\la^+[j]\over
\k})\}&(E6.2)\cr&\times\prod_{s_i}\{\cos(\pi{s_1\la^+[1]+\la^+[5]\over
\k})-\cos(\pi{\la^+[6]+s_2\la^+[2]+s_3\la^+[3]
+s'\la^+[4]\over \k})\}&}$$
where the second product is over all 8 choices of signs
$s_1,s_2,s_3\in\{\pm 1\}$, and $s'=s_1s_2s_3$.

\medskip\noindent{3.E7. {\it The algebra $E_7$.}}
The colabels for $E_7$ are 1,2,3,4,3,2,2, respectively.
The  scale factor in (1b) is $s=-\i/\sqrt{2}$, and $\k=k+18$.
To the weight $\la=(\la_1,\la_2,\ldots,\la_7)$ associate the orthogonal
coordinates $\la[\ell]={\la_7-\la_2\over 2}+\sum_{i=2}^\ell\la_i$ for
$\ell\le 6$, (interpret $\sum_{i=2}^1\la_i$ as 0) and $\la[7]=2\la_1+3\la_2+4\la_3+3\la_4+2\la_5+\la_6+2\la_7$.

The 63 coset representatives for $W(A_1D_6)\backslash W(E_7)$ are given in Table 1,
recursively defined in terms of the reflections $r_i$, $i=1,\ldots,7$,
 through the simple roots $\alpha_i$. Then
$$\eqalignno{S_{\la\mu}=&\,{32\sqrt{2}\over \k^{7/2}}\sum_{\ell=1}^{63}({\rm
det}\,c_\ell)\, \sin(\pi{(c_\ell.\la)^+[7]\,\mu^+[7]\over\k})&(E7.1)\cr
&\times\{{\rm det}(\cos(2\pi{(c_\ell.\la)^+[i]\,\mu^+[j]\over \k})_{1\le
i,j\le 6}-{\rm det}(\sin(2\pi{(c_\ell.\la)^+[i]\,\mu^+[j]\over
\k})_{1\le i,j\le 6})\}& \cr}$$
Recall that the numbers $(c_\ell.\la)^+[i]$ are the orthogonal
coordinates of the weight $c_\ell(\la+\rho)$.

The orbit representative $[\la]$ in (7c) can be found as follows.

\smallskip\item{(i)} Choose one of the 63 coset representatives $c_\ell$. Write
$x_i:=(c_\ell.\la)^+[7-i]$, $i=1,\ldots,6$, and $x_7:=(c_\ell.\la)^+[7]$,
for the orthogonal coordinates of $c_\ell(\la+\rho)$.
(The `$[7-i]$' is to make the indices here consistent
with those of $D_6$).

\item{(ii)} By adding an appropriate multiple of $2\k$ to $x_7$ and
replacing $x_7$ if necessary by $\k-x_7$, get
$0\le y_7<3\k/2$. Similarly, using the $D_6$ movements for the given
value of $\k$, get $0\le |y_6|<y_5<y_4<y_3<y_2<y_1$ and $y_1+y_2<\k$.

\item{(iii)} Write $t=(-y_1-y_2-y_3-y_4-y_5+y_6+y_7)/2$. If both $t>0$ and
$t\in\Z$, then the Dynkin labels of the desired weight
$[\la]$ are 
$$[\la]_1=t-1\ ,\ [\la]_7=y_5+y_6-1\ ,\ [\la]_i=x_{7-i}-x_{8-i}-1\qquad 2\le
i\le 6$$

\item{(iv)} Otherwise replace $y_7=\k-y_7$, $y_1=y_6-\k/2$,
$y_6=y_1+\k/2$, and $y_i=\k/2-y_{7-i}$ for $i=2,3,4,5$. Define $t$ as
in (iii); if both $t>0$ and $t\in\Z$ then write $[\la]_i$ as in (iii).

\item{(v)} Otherwise try the next $c_\ell$ in (i).\smallskip

Equation (8) becomes 
$$ \eqalignno{\epsilon(\la)={\rm sign}\, &\sin(\pi{\la^+[7]\over \k})\prod_{1\le i<j\le 6}
\{\cos(2\pi{\la^+[i]\over
\k})-\cos(2\pi{\la^+[j]\over\k})\}&(E7.2)\cr\times&\prod_{s_i}
\{\cos(\pi{s_1\la^+[1]+s_2\la^+[2]+s_3\la^+[3]+\la^+[4]\over
\k})-\cos(\pi{\la^+[7]+s_5\la^+[5]
-s'\la^+[6]\over \k})\}&}$$
where the second product is over all 16 choices of signs $s_1,s_2,s_3,s_5\in\{\pm 1\}$,
and $s'=s_1s_2s_3s_5$.

\medskip\noindent{3.E8. {\it The algebra $E_8$.}}
The colabels here are 2,3,4,5,6,4,2,3, respectively. The scale factor
in (1b) is $s=1$, and $\k=k+30$.
To the weight $\la=(\la_1,\la_2,\ldots,\la_8)$ associate the orthogonal
coordinates $\la[\ell]={\la_8-\la_6\over 2}+\sum_{i=8-\ell}^6\la_i$
for $1\le \ell\le 7$, (interpret $\sum_{i=7}^6\la_i$ as 0) and
$\la[8]=\la_1+2\la_2+3\la_3+4\la_4+5\la_5+{3\over
2}\la_6+2\la_7+{5\over 2}\la_8$.

The 135 coset representatives for $W(D_8)\backslash W(E_8)$ are given in Table 2,
recursively defined in terms of the reflections $r_i$, $i=1,\ldots,8$,
 through the simple roots $\alpha_i$. Then
$$\eqalignno{S_{\la,\mu}={128\over \k^4}\sum_{\ell=1}^{135}({\rm
det}\,c_\ell)&\,\{{\rm
det}(\cos(2\pi{(c_\ell.\la)^+[i]\,\mu^+[j]\over  \k}))_{1\le i,j\le
8}&(E8.1)\cr &+{\rm det} (\sin(2\pi{(c_\ell.\la)^+[i]\,\mu^+[j]\over
\k}))_{1\le i,j\le 8}\}& \cr}$$
Recall that the numbers $(c_\ell.\la)^+[i]$ are the orthogonal
coordinates of the weight $c_\ell(\la+\rho)$.

The orbit representative $[\la]$ in (7c) can be found as follows.

\smallskip\item{(i)} Choose one of the 135 coset representatives $c_\ell$. Write
$x_i:=(c_\ell.\la)^+[9-i]$, $i=1,\ldots,8$, for the orthogonal coordinates of
$c_\ell(\la+\rho)$ (the `$[9-i]$' is to make the indices here consistent
with those of $D_8$).

\item{(ii)} Using the $D_8$ movements for the given
value of $\k$, get $0\le |y_8|<y_7<\cdots<y_1$ and $y_1+y_2<\k$.

\item{(iii)} Write $t=y_1-y_2-y_3-y_4-y_5-y_6-y_7+y_8$. If $t>0$
 then the Dynkin labels of the desired weight
$[\la]$ are 
$$[\la]_7=t/2-1\ ,\ [\la]_8=y_7+y_8-1\ ,\ [\la]_i=y_{i+1}-y_{i+2}-1\qquad 1\le
i\le 6$$

\item{(iv)} Otherwise replace each $y_i$ with $\k/2-y_{9-i}$. Compute
the resulting $t$ as in (iii); if $t$ is positive than write $[\la]_i$ as in (iii).

\item{(v)} Otherwise try the next coset representative $c_\ell$ in (i).\smallskip

Equation (8) becomes 
$$ \eqalignno{\epsilon(\la)=&\,{\rm sign} \prod_{1\le i<j\le 8}
\{\cos(2\pi{\la^+[i]\over
\k})-\cos(2\pi{\la^+[j]\over\k})\}&(E8.2)\cr\times&\prod_{s_i}
\{\cos(\pi{s_1\la^+[1]+s_2\la^+[2]+s_3\la^+[3]+\la^+[7]\over
\k})-\cos(\pi{s_4\la^+[4]+s_5\la^+[5]
+s'\la^+[6]+\la^+[8]\over \k})\}&}$$
where the second product is over all 32 choices of signs
$s_1,s_2,s_3,s_4,s_5\in\{\pm 1\}$, and where $s'=s_1s_2s_3s_4s_5$.

\medskip\noindent{3.F4. {\it The algebra $F_4$.}}
The colabels are 2,3,2,1, resp. The scale factor in (1b)
is $s=1/2$ and $\k=k+9$.
To the weight $\la=(\la_1,\la_2,\la_3,\la_4)$ associate the orthogonal
coordinates $$(\la[1],\la[2],\la[3],\la[4])=
(\la_1+2\la_2+{3\over 2}\la_3+\la_4,\la_1+\la_2+{1\over 2}\la_3,\la_2+{1\over 2}\la_3,
{1\over 2}\la_3)$$

Let $r_i$, $i=1,\ldots,4$ be the reflections through the simple roots
$\alpha_i$. Define the matrix 
$$c={1\over 2}\left(\matrix{1&1&1&-1\cr 1&1&-1&1\cr 1&-1&1&1\cr 1&-1&-1&-1}\right)$$
It is orthogonal, with determinant $+1$. Then
$$\eqalignno{S_{\la,\mu}={8\over \k^2}\{&{\rm
det}(\sin(2\pi{\la^+[i]\,\mu^+[j]\over \k}))_{1\le i,j\le 4}+{\rm 
det}(\sin(2\pi{(c.\la)^+[i]\,\mu^+[j]\over \k}))_{1\le i,j\le 4}&\cr&+{\rm 
det}(\sin(2\pi{(c^t.\la)^+[i]\,\mu^+[j]\over \k}))_{1\le i,j\le 4}\}
&(F4.1)}$$
In (F4.1), $(c.\la)^+[i]$ denotes the $i$th orthogonal coordinate of
the matrix product of $c$ with the column vector with entries
$(\la+\rho)[j]$, and $c^t$ means the transpose(=inverse) of $c$.

The orbit representative $[\la]$ in (7c) can be found as follows. 

\smallskip\item{(i)} Let
$x_i=\la^+[i]$ for $i=1,2,3,4$. Apply the $B_4$ algorithm (steps
(ii)--(iv)) with the
given value of $\k$ to obtain the 4-tuple $(y_i)$, $\k>y_1>y_2>y_3>y_4>0$. 

\item{(ii)} If $y_1>y_2+y_3+y_4$, proceed directly to (iii); otherwise
first replace $y_i$ with the coordinates
of the matrix product of $c^t$ with the column vector $(y_i)$, and if
necessary change the sign of the resulting $y_4$ so that it's positive.

\item{(iii)} The desired weight $[\la]$ has Dynkin labels
$$[\la]_1=y_2-y_3-1\ ,\ [\la]_2=y_3-y_4-1\ ,\ [\la]_3=2y_4-1\ ,\ [\la]_4=y_1-y_2-y_3-y_4-1$$ 

Equation (8) becomes 
$$ \eqalignno{\epsilon(\la)={\rm sign} &\prod_{i=1}^4\sin(\pi{\la^+[i]\over \k})
\prod_{1\le i<j\le 4}\{\cos(2\pi{\la^+[j]\over
\k})-\cos(2\pi{\la^+[i]\over\k})\}&\cr&\prod_{s_2,s_3} \{\cos(\pi{\la^+[4]\over
\k})-\cos(\pi{\la^+[1]+s_2\la^+[2]+s_3\la^+[3]\over \k})\}&(F4.2)}$$
where the product is over the four choices of signs $s_2,s_3\in\{\pm 1\}$.

The embedding $B_4\subset F_4$ (resp.\ $D_4\subset F_4$) lets us write
the $F_{4,k}$ quantities in terms of the more familiar $B_{4,k+2}$
(resp.\ $D_{4,k+3}$) ones, and this can be useful (see [31]). In
particular, a weight $\la\in P_+^k(F_4)$ corresponds to the orbit of
$\la'=(\la_2+\la_3+\la_4+2,\la_1,\la_2,\la_2+\la_3+1)\in
P_+^{k+1}(D_4)$, under $P':\la'_3\leftrightarrow \la_4'$ (which
changes the chirality of $D_4$ weights) and triality
$T':\la_1\rightarrow\la_3\rightarrow\la_4\rightarrow\la_1$ (we prime
the $D_4$ quantities). We get
$S_{\la\mu}=\sum_{i=0}^1\sum_{j=0}^2(-1)^iS'_{P'{}^iT'{}^j\la',\mu'}$
and 
$$N_{\la\mu}^\nu=\sum_{\gamma'}\sum_{i=0}^1\sum_{j=0}^2(-1)^i\,b_{\gamma'}^\la\,N'_{\gamma',P'{}^i
T'{}^j\mu'}{}^{\nu'}$$
where $\oplus_{\gamma'}\,b_{\gamma'}^\la\,(\gamma')=(\la)$ are $D_4\subset F_4$ branching
rules (see e.g.\ [30]).

\medskip\noindent{3.G2. {\it The algebra $G_2$.}}
The colabels here are 2,1, respectively. The scale factor in (1b) is
$s=-1/\sqrt{3}$, and $\k=k+4$.
To the weight $\la=(\la_1,\la_2)$ associate the orthogonal
coordinates $(\la[1],\la[2],\la[3])=(\la_1+\la_2,\la_1,-2\la_1-\la_2)$.
The finite Weyl group is
the symmetric group $S_{3}$, together with the sign change
$\la\mapsto-\la$.  Then
$$\eqalignno{S_{\la,\mu}={-2\over\sqrt{3}\k}&
\{c({2mm'\!+\!mn'\!+\!nm'\!+\!2nn'})+c({-mm'\!-\!2mn'\!-\!nn'\!+\!nm'})&(G2.1)\cr&+
c({-mm'+mn'-2nm'-nn'})-c({-mm'-2mn'-2nm'-nn'})&\cr&-c({2mm'+mn'+nm'-nn'
})-c({-mm'+mn'+nm'+2nn'})\}&}$$
where for convenience we put $c(x)=\cos(2\pi x/3\k)$,
$m=\la^+[1]$, $n=\la^+[2]$, $m'=\mu^+[1]$,
and $n'=\mu^+[2]$.

The orbit representative $[\la]$ in (7c) can be found as follows. 

\smallskip\item{(i)} Let $x_i=\la^+[i]$.

\item{(ii)} By adding the appropriate multiples of $\k$ to each $x_i$, find the
unique numbers $0<y_1,y_2,y_3<\k$ such that $x_i\equiv y_i$ (mod $\k$).

\item{(iii)} If $y_1+y_2+y_3=2\k$, then replace each $y_i$ with $\k-y_i$.

\item{(iv)} Reorder the $y_i$ so that $y_3>y_1>y_2$.
 The Dynkin labels of the desired weight are $[\la]_1=y_2-1$
and $[\la]_2=y_1-y_2-1$.

\smallskip Equation (8) becomes 
$$ \epsilon(\la)={\rm sign}\{s(3a)\,s(b)\,s(3a+b)\, s(3a+2b)\,s(a+b)\,s(
6a+3b)\}\eqno(G2.2)$$
where for convenience we write $a=\la_1+1$ and $b=\la_2+1$, and put
$s(x)=\sin(\pi x/3\k)$.

The embedding $A_2\subset G_2$  lets us write
the $G_{2,k}$ quantities in terms of the more familiar $A_{2,k+1}$
 ones, and this can be useful (see [31]). In
particular, any weight $\la\in P_+^k(G_2)$ corresponds to a pair
$\la'=(\la_1,\la_1+\la_2+1),C'\la'=(\la_1+\la_2+1,\la_1)$ of weights in
$P_+^{k+1}(A_2)$. We get
$S_{\la\mu}=\i\,S'_{\la'\mu'}-\i\,\overline{S'_{\la'\mu'}}$
and 
$$N_{\la\mu}^\nu=\sum_{\gamma'}\,b_{\gamma'}^\la\,N'_{\gamma'\mu'}{}^{\nu'}
-N'_{\gamma',C'\mu'}{}^{\nu'}$$
where $\oplus_{\gamma'}\,b_{\gamma'}^\la\,(\gamma')=(\la)$ are $A_2\subset G_2$ branching
rules (see e.g.\ [30]), and $C':\mu'_1\leftrightarrow\mu'_2$ is $A_2$ 
charge-conjugation.

\bigskip\bigskip\noindent{{\bf 4. Further remarks}}\bigskip 

\noindent{{\it 4.1. Error analysis and exact results.}} The error
analysis of the Gaussian elimination method for computing determinants
is surprisingly subtle and has been the subject of extensive study ---
see e.g.\ [22]. It depends on the size of the pivots, but very
typically the error grows by a factor on the order of $\sqrt{n}$ for
an $n\times n$ matrix. A good idea however is to make sure in your
program that the pivots are never too close to 0.

Because our quantities $S_{\la\mu}$ are all cyclotomic integers (up
to a global rescaling) in some field $\Q[\exp[2\pi\i/n]]$, the obvious
way to make all calculations exact is to do it over the polynomials
$p(x)$ with integer coefficients. The desired numerical value would then
simply be $p(\exp[2\pi\i/n])$. These polynomials add, multiply etc in
the usual way, but we regard two polynomials as equal if they differ
by a multiple of the cyclotomic polynomial $\phi_n(x)$. The cyclotomic
polynomial is the polynomial of smallest positive degree, with integer
coefficients, which has $\exp[2\pi\i/n]$ as a root. For example,
$\phi_p(x)=x^{p-1}+x^{p-2}+\cdots+1$ and
$\phi_{n}(x)=\phi_{p_1\cdots p_t}(x^{n/p_1\cdots p_t})$ where
$n=p_1^{e_1}\cdots p_t^{e_t}$. A manifestly integral algorithm for
computing any $\phi_n$ is provided in [25]. 
Because $\phi_n$ has
degree $\varphi(n)=\prod_i(p_i^{e_i}-p_i^{e_i-1})$, we can require
each of our polynomials $p(x)$ to be of degree less than $\varphi(n)$.

But there is an alternate approach, which is perhaps a little
simpler. We represent each number by an integer polynomial
$p(x)$, and always reduce any exponents modulo $n$ (i.e.\ identify $x^n$
and 1, $x^{n+1}$ and $x$, etc). When $n$ is even, we can if we like
reduce exponents modulo $n/2$, by identifying $x^{n/2}$ with $-1$, etc.
We equate two polynomials
$p(x),q(x)\in\Z[x]$ if $$|p(\exp[2\pi\i k/n])-q(\exp[2\pi\i k/n])|<0.5$$
holds for all $k$, $1\le k<n$, coprime to $n$. Of course, `$(\exp[2\pi\i k/n])^m$' is
evaluated as `$\cos(2\pi km/n)+\i\sin(2\pi km/n)$, to decrease
error. The point is that if an algebraic integer $z$ (such as 
$z=p(\exp[2\pi\i k/n])-q(\exp[2\pi\i k/n])$) and all of its Galois
associates $\si(z)$ have modulus $<1$, then $z$ must equal 0. The
reason is that $\prod_\si(-\si(z))\in\Z$ is the constant term in the
minimal polynomial of $z$.

\medskip\noindent{{\it 4.2. Character formulas.}} The expressions in \S3 for
$S$ don't depend on the integrality of the weight $\mu$, and thus
taking the appropriate ratio (4a) gives determinant expressions for
finite-dimensional Lie group characters.

\medskip\noindent{{\it 4.3. Fusion coefficients.}} 
Recall the Kac-Walton formula (2c). Consider any $\la,\mu\in P_+^k$.
A tensor product coefficient
$T_{\la\mu}^\nu$ will contribute $\epsilon(\nu)\,T_{\la\mu}^{\nu}$
to the fusion coefficient $N_{\la\mu}^{[\nu]}$ (recall that $\epsilon(\nu)$ and $[\nu]$
are computed in \S3). In particular,
 any null weight $\nu$ in the tensor product of $\la$ and $\mu$
 can be ignored. When the level is sufficiently high (specifically, $\la_0
 +\mu_0\ge k$),
 the fusion product of $\la$ and $\mu$ will equal their tensor product.


\medskip\noindent{{\it 4.4. Galois action.}} By (1c) we see that the
entries of $S$ lie in
a cyclotomic field $\Q[\xi_N]$ for some root of unity $\xi_N=\exp[2\pi\i/N]$.
This simply means that each $S_{\la\mu}$
can be written as a polynomial $p_{\la\mu}(\xi_N)$ in $\xi_N$ with rational coefficients.
For $A_r,B_r,\ldots,G_2$, respectively, we can take
$$N=4\,(r+1)\k,4\k,4\k,4\k,12\k,8\k,2\k,2\k,6\k$$
but usually this is larger than necessary (see [18]).

Take any Galois automorphism $\si\in{\rm Gal}(\Q[\xi_N]/\Q)$. 
The group Gal$(\Q[\xi_N]/\Q)$ is isomorphic to the multiplicative group
$(\Z/N\Z)^\times$ of numbers coprime to $N$, so to $\si$ we can assign an
integer $\ell$; explicitly, $\si=\si_\ell$ takes the number $p_{\la\mu}(\xi_N)$
to $p_{\la\mu}(\xi_N^\ell)$. For example, $\ell=-1$ corresponds to complex
conjugation. To the Galois automorphism $\si$ is a sign
$\epsilon_\si:P_+^k\rightarrow \{\pm\}$ and a permutation $\la\mapsto
\si(\la)$ of $P_+^k$, such that [16]
$$\si(S_{\la\mu})=\epsilon_\si(\la)\,S_{\si(\la),\mu}=\epsilon_\si(\mu)\,
S_{\la,\si(\mu)}\eqno(9a)$$
For example, $\ell=-1$ corresponds to `charge-conjugation' $\la\mapsto C\la$
and parity $\epsilon_{-1}(\la)=+1$. Some properties and formulas for
$\epsilon_\si(\la)$ for $A_{r,k}$ are given in [20].
For arbitrary $\si=\si_\ell$,
the permutation $\la\mapsto\si(\la)$ and parity $\epsilon_\si(\la)$
are computed by
$$\eqalignno{\si(\la)=&\,[\ell(\la+\rho)-\rho]&(9b)\cr
\epsilon_\si(\la)=&\,\epsilon'_\ell\,\epsilon(\ell(\la+\rho)-\rho)&(9c)}
$$

A major motivation for writing this paper was to make this Galois
action more accessible. In particular, 
efficient and explicit algorithms for computing $[\nu]$ were given in \S3 for each
algebra, and by far the fastest algorithms known to this author for computing
$\epsilon(\nu)$ use the simplifications to (8) given explicitly in
each subsection of \S3. The sign $\epsilon'_\ell$ is a number-theoretic parity related to
quadratic residues, and rarely is needed in the applications as it is
independent of the weights. It can be computed as follows [16]:

\smallskip Write $\k^{r/2}s$ in the form $R\sqrt{M}$ where $R\in\Q$ and $M\in\Z$ is
square-free. Then $\epsilon'_\ell$ equals the Jacobi symbol $({M\over\ell})\in
\{\pm 1\}$. This is quickly computed, using standard properties of the Jacobi symbol
(see e.g.\ [26]), e.g.\ quadratic reciprocity and factorisation.\smallskip

For example, consider again $A_{3,6}$. Here, $N=40$
works, so for the representatives $\ell\in(\Z/40\Z)^\times$ we can take
$\ell=\pm 1,\pm 3,\pm 7,\pm 9,\pm 11,\pm 17,\pm 19$. Then, respectively, 
$$\eqalign{\si_\ell(0)=&\,0,(2,2,2),(2,0,2),(0,6,0),(0,6,0),(2,2,2),(0,0,0)\cr
\epsilon(\ell\rho-\rho)=&\, 1,1,1,1,1,1,1\cr
\epsilon'_\ell=&\,1,1,-1,1,-1,-1,- 1}$$

\medskip\noindent{{\it 4.5. Modular invariants.}} The 1-loop partition
function of WZW CFT looks like
$${\cal Z}=\sum_{\la,\mu\in P_+^k}M_{\la\mu}\,\chi_\la\,\chi_\mu^*$$
where (among other things) each $M_{\la\mu}\in\Z_{\ge}$ and $MS=SM$.
Hitting $MS=SM$ with a Galois automorphism $\si$ implies from (9a) that
$$\eqalignno{M_{\si \la,\si \mu}=&\,M_{\la\mu}&(10a)\cr
M_{\la\mu}\ne 0\quad\Rightarrow&\quad \epsilon_\si(\la)=\eps_\si(\mu)&(10b)}$$
valid for all $\si$. The `parity rule' (10b) turns out to be especially
powerful. Efficient  algorithms to compute $S$ entries and
parities are vital in this context and was the primary motivation for
writing this paper. See [6] for more details and examples.

\medskip\noindent{{\it 4.6. Branching rules.}} When $X_r\subset \widetilde{X}_s$,
and the central charge $c$ of $X_{r,k}$ equals that of $\widetilde{X}_{s,\ell}$,
then we say that we have a {\it conformal embedding}. In this case, the
level $\ell$ modules $\widetilde{L(\la)}$ of $\widetilde{X}_s^{(1)}$ can
be decomposed into a finite direct sum $\oplus_iL(\la^{(i)})$ of level $k$ modules
of $X_r^{(1)}$. These decompositions are called {branching rules}.

All conformal embeddings are known [27], and most of their branching rules are
known (see e.g.\ [28] and references therein), but some branching rules don't seem to appear explicitly in the
literature, or only appear conjecturally.

Branching rules can be read off from the appropriate modular invariants
(see e.g.\ [29]).
%
In [6] we use the previous algorithms and formulae
to obtain branching rules for the conformal
embeddings. Mostly this merely
provides an independent check; however it also fills some gaps in the
literature.

\bigskip
\noindent{{\bf Acknowledgements}}

I warmly thank Mark
Walton, who over the years has been my sounding board for this and many related questions.
I also thank Christoph Schweigert for informing me that [17] used determinant
formulas for the classical algebras. Matthias Gaberdiel helped me spot
the error in (G2.1) in an earlier draft of this paper.
This research was supported in part by NSERC. It was written in part
at St.\ John's College, Cambridge, who I thank for their considerable
hospitality. 

\bigskip\centerline{{\bf References}}
\medskip

\item{1.} {Fuchs, J.\ and  van Driel}, P.: 
{\it Nucl.\ Phys.} {\bf B346} (1990), 632--648;

\item{} {\it ibid.}: 
{\it Lett.\ Math.\ Phys.} {\bf 23} (1991), 11--18.

\item{2.} Fuchs, J.: In: {\it Sakharov Memorial Lectures in Physics},
Nova Sci.\ Publ., Commack, NY, 1992.

\item{3.} Moody, R.\ V.\ and Patera, J.: {\it Siam J.\ Alg.\ Discr.\
Meth.} {\bf 5} (1984), 359--383;

\item{} {\it ibid.}: 
{\it Math.\ of Comput.} {\bf 48} (1987), 799--827.

\item{4.} Karadayi, H.\ R.\ and Gungormez, M.: {\it J.\ Phys.} {\bf A32}
(1999), 1701--1707;

\item{} {\it ibid.}: ``Summing over the Weyl groups of $E_7$ and $E_8$'',
math-ph/9812014;

\item{} {\it ibid.}: ``On the calculation of group characters'', math.ph/0008014.

\item{5.} {Kac, V.\ G.\ and Peterson}, D.\ H.: {\it Adv.\ Math.} {\bf 53}
(1984), 125--264.

\item{6.} Gannon, T.: ``The modular invariants of the exceptional algebras'',
{\it in preparation};

\item{} {\it ibid.}: ``The modular invariants of the low rank classical
algebras'', {\it work in progress}.

\item{7.} Verlinde, E.: {\it Nucl.\ Phys.} {\bf B300} (1988), 360--376.

\item{8.} {Kac, V.\ G.}: {\it Infinite Dimensional Lie algebras},
3rd edn, Cambridge University Press, Cambridge, 1990.

\item{9.} {Walton}, M.\ A.: 
{\it Phys.\ Lett.} {\bf B241} (1990), 365--368.

\item{10.} Bourbaki, N.: {\it Groupes et Alg\'ebres de Lie, Chapitres IV-VI},
Paris, Hermann, 1968.

\item{11.} Kass, S., Moody, R.\ V., Patera, J., and Slansky, R.,
{\it Affine Lie Algebras, Weight Multiplicities, and Branching Rules}, Vol.\ 1
(Univ.\ of California Press, Berkeley, 1990).

\item{12.} {Fuchs, J.\ and Schweigert}, C.: {\it
Symmetries, Lie algebras, and Representations}, Cambridge Univ.\
Press, Cambridge, 1997.

\item{13.} {Di Francesco, Ph., Mathieu,} P.\  and 
S\'en\'echal, D.: {\it Conformal Field Theory}, Springer, New York, 1997.

\item{14.} Gannon, T.: ``Modular data: the algebraic combinatorics of
conformal field theory'', math.QA/0103044.

\item{15.} Bremner, M.\ R., Moody, R.\ V., and Patera, J.: {\it Tables of
Dominant Weight Multiplicities for Representations of Simple Lie Algebras}
(Marcel Dekkar, New York, 1985).

\item{16.} {Coste, A.} and {Gannon, T.}: 
{\it Phys.\ Lett.} {\bf B323} (1994), 316--321.

\item{17.} Mlawer, E.\ J., Naculich, S.\ G., Riggs, H.\ A., and Schnitzer,
H.\ J.: {\it Nucl.\ Phys.} {\bf B352} (1991), 863--896.

\item{18.} Gannon, T.\ and Walton, M.\ A.: {\it Commun.\ Math.\ Phys.}
{\bf 206} (1999), 1--22.

\item{19.} Idowu, A.\ J.\ and Morris, A.\ O.: 
{\it Math.\ Proc.\ Camb.\ Phil.\ Soc.} {\bf 101} (1987), 405--420.

\item{20.} Altsch\"uler, D., Ruelle, P.\ and Thiran, E.: {\it J.\
Phys.} {\bf A32} (1999), 3555--3570. 

\item{21.} Knuth, D.E.: {\it The Art of Computer Programming}, Vol.\ 3,
2nd ed., Addison-Wesley (Reading, 1998).

\item{22.} Trefethen, L.N.\ and Bau, D.: {\it Numerical Linear Algebra}
(SIAM, Philadelphia, 1997).

\item{23.} Fulton, W.\ and Harris, J.: {\it Representation Theory: A
first course}, Springer, New York, 1991.

\item{24.} {Ruelle, P., Thiran, E.\ and Weyers}, J.: 
{\it Nucl.\ Phys.} {\bf B402} (1993), 693--708.

\item{25.} Bloom, D.M.: {\it Amer.\ Math.\ Monthly} {\bf 75} (1968), 372--377.

\item{26.} Hua Loo Keng, {\it Introduction to Number Theory},
Springer, Berlin, 1982.

\item{27.} Bais, F.\ A.\ and  Bouwknegt, P.\ G.: {\it Nucl.\ Phys.} {\bf B279}
(1987), 561--570;

\item{} Schellekens, A.\ N.\ and Warner, N.\ P.: {\it Phys.\ Rev.} {\bf D34}
 (1986), 3092--3096;

\item{} Arcuri, R.C., Gomes, J.F., and Olive, D.I.: {\it Nucl.\ Phys.}
{\bf B285} (1987), 327--339. 

\item{28.}  {Kac, V.\ G.\ and Wakimoto}, M.: 
{\it Adv.\ Math.} {\bf 70} (1988), 156--236;
\item{} Kac, V.G.\ and Sanielevici, M.: {\it Phys.\ Rev.} {\bf D37} 2231--2237
(1988);
\item{} Walton, M.A.: {\it Nucl.\ Phys.} {\bf B322} 775--790 (1989);

\item{} Verstegen, D.: {\it Commun.\ Math.\ Phys.} {\bf 137} 567--586 (1991); 

\item{}
Levstein, F.\ and Liberati, J.I.: {\it Commun.\ Math.\ Phys.}
{\bf 173} (1995), 1--16;

\item{} Schellekens, A.N.: in {\it Proc.\ of XXVIII Intern.\ Symp.\
Ahrenshoop on Theory of Elementary Particles}, D.\ L\"ust and G.\
Weigt (eds.), DESY publ., Zeuthen, 1995, 1.

\item{29.} Bouwknegt, P.\ and Nahm, W.: {\it Phys.\ Lett.} {\bf B184}
(1987),  359--362.

\item{30.} McKay, W.G.\ and Patera, J.: {\it Tables of Dimensions,
Indices, and Branching Rules for Representations of Simple Lie
Algebras}, Marcel Dekkar, Inc., New York, 1981.

\item{31.} Gaberdiel, M.R.\ and Gannon, T.: ``Boundary states for WZW
models'', hep-th/0202067.

\vfill\eject
\centerline {Table 1. Coset representatives for $W(D_5)\backslash W(E_6)$ and
$W(A_1D_6)\backslash W(E_7)$}
\medskip
$$\vbox{\tabskip=0pt\offinterlineskip
  \def\tablerule{\noalign{\hrule}}
  \halign to 5.75in{
    \strut#&\vrule#\tabskip=0em plus1em &    
    \hfil#&\vrule#&\hfil#&\vrule#&    
    \hfil#&\vrule#\tabskip=0pt\cr\tablerule     
&&$\,$length$\,$&&
 \hidewidth $c_i$ for $E_6$\hidewidth&&\hidewidth $c_i$\ {\rm for}\ $E_7$\hidewidth&\cr
\tablerule&&\hfil 0\hfil && \hfil$c_1=I$\hfil &&\hfil$c_1=I$\hfil&\cr
&& \hfil 1\hfil && \hfil$c_2=r_1$\hfil &&\hfil$c_2=r_1$\hfil&\cr
&&\hfil 2\hfil && \hfil$c_3=c_2r_2$\hfil &&\hfil$c_3=c_2r_2$\hfil&\cr
&& \hfil 3\hfil&& \hfil$c_4=c_3 r_3$\hfil &&\hfil$c_4=c_3r_3$\hfil&\cr
&&\hfil 4\hfil && \hfil$c_5=c_4r_4\ ,\ c_6=c_4r_6$\hfil &&\hfil$c_5=c_4r_4\ ,\ c_6=c_4r_7$\hfil&\cr
&&\hfil 5\hfil && \hfil$c_7=c_5r_5\ ,\ c_8=c_5r_6 $\hfil&&\hfil$c_7=c_5r_5\ ,\ c_8=c_5r_7$\hfil&\cr
&&\hfil 6\hfil &&\hfil $c_9=c_7r_6\ ,\ c_{10}=c_8r_3 $\hfil&&\hfil$c_9=c_7r_6\ ,\ c_{10}=c_7r_7\ ,\
c_{11}=c_8r_3$\hfil&\cr
&&\hfil 7\hfil && \hfil$c_{11}=c_9r_3\ ,\ c_{12}=c_{10}r_2$\hfil&&\hfil$c_{12}=c_9r_7\ ,\ c_{13}=c_{10}r_3\ ,\
 c_{14}=c_{11}r_2 $\hfil&\cr
&&\hfil 8\hfil && \hfil$c_{13}=c_{11}r_2\ ,\ c_{14}=c_{11}r_4$ ,\hfil&&
\hfil$c_{15}=c_{12}r_3\ ,\ c_{16}=c_{13}r_2\ ,\ c_{17}=c_{13}r_4\ ,\ c_{18}=c_{14}r_1$\hfil&\cr&&&&\hfil$c_{15}=c_{12}r_1$\hfil&&&\cr
&&\hfil 9\hfil && \hfil$c_{16}=c_{13}r_1\ ,\ c_{17}=c_{13}r_4$\hfil&&\hfil
$c_{19}=c_{15} r_2\ ,\ c_{20}=c_{15}r_4\ ,\ c_{21}=c_{16}r_1\ ,\ c_{22}=c_{16}r_4$\hfil&\cr
&&\hfil 10\hfil && \hfil$c_{18}=c_{16}r_4\ ,\ c_{19}=c_{17}r_3$\hfil&&\hfil
$c_{23}=c_{19}r_1\ ,\ c_{24}=c_{19}r_4\ ,\ c_{25}=c_{20}r_5$
,\hfil&\cr&&&&&&\hfil$  c_{26}=c_{21}r_4\ ,\ c_{27}=c_{22}r_3$\hfil &\cr
&&\hfil 11\hfil &&\hfil $c_{20}=c_{18}r_3\ , c_{21}=c_{19}r_6$\hfil&& \hfil
$c_{28}=c_{23}r_4\ ,\ c_{29}=c_{24}r_3\ ,\ c_{30}=c_{24}r_5$ ,\hfil&\cr&&&&&&\hfil$
c_{31}=c_{26}r_3\ 
,\ c_{32}=c_{27}r_7$\hfil&\cr&&\hfil 12\hfil && \hfil$c_{22}=c_{20}r_2\ , c_{23}=c_{20}r_6$\hfil&&\hfil
$c_{33}=c_{28}r_3\ ,\ c_{34}=c_{28}r_5\ ,\ c_{35}=c_{29}r_5$ ,\hfil&\cr&&&&&&
 \hfil$c_{36}=c_{29}
r_7\ ,\ c_{37}=c_{31}r_2\ ,\ c_{38}=c_{31} r_7 $\hfil&\cr&&\hfil 13\hfil &&\hfil$ c_{24}=c_{22}r_6$\hfil&&\hfil
$c_{39}=c_{33} r_2\ ,\ c_{40}=c_{33} r_5\ ,\ c_{41}=c_{33} r_7$ ,\hfil
&\cr&&&&&&\hfil$ c_{42}=
c_{35} r_4\ ,\ c_{43}=c_{35}r_7\ ,\ c_{44}=c_{37}r_7$\hfil&\cr&&\hfil 14\hfil &&\hfil$c_{25}=c_{24}r_3$\hfil&&\hfil
$c_{45}=c_{39} r_5\ ,\ c_{46}=c_{39}r_7\ ,\ c_{47}=c_{40}r_4$
,\hfil&\cr&&&&&&\hfil$ c_{48}=c_{40}
r_7\ ,\ c_{49}=c_{42}r_7\ ,\ c_{50}=c_{44} r_3$\hfil &\cr&&\hfil 15\hfil &&\hfil$c_{26}=c_{25}r_4$\hfil&&\hfil
$c_{51}=c_{45}r_4\ ,\ c_{52}=c_{45}r_7\ ,\ c_{53}=c_{46}r_3$ ,\hfil&\cr&&&&&&
\hfil$ c_{54}=c_{47}
r_7\ ,\ c_{55}=c_{49}r_3\ ,\ c_{56}=c_{50}r_4$\hfil&\cr&&\hfil 16\hfil &&\hfil$c_{27}=c_{26}r_5$\hfil&&\hfil
$c_{57}=c_{51} r_3\ ,\ c_{58}=c_{51}r_7\ ,\ c_{59}=c_{52}r_3\ ,\
c_{60}=c_{53}r_4$ ,\hfil&\cr&&&&&&\hfil$
c_{61}=c_{54}r_3\ ,\ c_{62}=c_{55}r_2\ ,\ c_{63}=c_{56} r_5 $\hfil&\cr
\tablerule\noalign{\smallskip}
 }} $$
\centerline{The coset representatives $c_j$ are given recursively in terms of
the simple reflections $r_i$.}\vfill\eject

\centerline {Table 2. Coset representatives for $W(D_8)\backslash W(E_8)$}
\medskip
$$\vbox{\tabskip=0pt\offinterlineskip
  \def\tablerule{\noalign{\hrule}}
  \halign to 5.75in{
    \strut#&\vrule#\tabskip=0em plus1em &    
    \hfil#&\vrule#&\hfil#&\vrule#\tabskip=0pt\cr\tablerule     
&&$\,$length$\,$&&
 \hidewidth$c_i$ for $E_8$\hidewidth&\cr\tablerule
&&\hfil 0\hfil &&\hfil $c_1=I$\hfil & \cr
&& \hfil 1\hfil && \hfil$c_2=r_7$\hfil & \cr
&&\hfil 2\hfil &&\hfil $c_3=c_2 r_6$\hfil & \cr
&& \hfil 3\hfil &&\hfil $c_4=c_3 r_5$\hfil & \cr
&&\hfil 4\hfil &&\hfil $c_5=c_4 r_4\ ,\ c_6=c_4 r_8$\hfil & \cr
&&\hfil 5\hfil &&\hfil $c_7=c_5 r_3\ ,\ c_8=c_5 r_8$\hfil & \cr
&&\hfil 6\hfil &&\hfil $c_9=c_7 r_2\ ,\ c_{10}=c_7 r_8\ ,\ c_{11}=c_8 r_5$\hfil & \cr
&&\hfil 7\hfil &&\hfil$c_{12}=c_9r_1\ ,\ c_{13}=c_9r_8\ ,\ c_{14}=c_{10}r_5\ ,\ c_{15}=c_{11}r_6$\hfil&\cr
&&\hfil 8\hfil &&\hfil$c_{16}=c_{12}r_8\ ,\ c_{17}=c_{13}r_5\ ,\ c_{18}=c_{14}r_4\ ,\
c_{19}=c_{14}r_6\ ,\ c_{20}=c_{15} r_7$\hfil&\cr&&\hfil 9\hfil &&
\hfil$c_{21}=c_{16} r_5\ ,\ c_{22}=c_{17}r_4\ ,\ c_{23}=c_{17}r_6\ ,\ c_{24}=c_{18}
 r_6\ ,\ c_{25}=c_{19} r_7$\hfil&\cr&& \hfil 10\hfil &&\hfil
$c_{26}=c_{21} r_4\ ,\ c_{27}=c_{21}r_6\ ,\ c_{28}=c_{22}r_3\ ,\ c_{29}=c_{22}
r_6$ ,\hfil&\cr&&&&\hfil$ c_{30}=c_{23}r_7\ ,\ c_{31}=c_{24}r_5\ ,\
c_{32}=c_{24}r_7$\hfil&\cr&&\hfil 11\hfil&&\hfil
$c_{33}=c_{26}r_3\ ,\ c_{34}=c_{26}r_6\ ,\ c_{35}=c_{27}r_7\ ,\ c_{36}=c_{28}
r_6$ ,\hfil&\cr &&&&\hfil$c_{37}=c_{29}r_5\ ,\ c_{38}=c_{29}r_7\ ,\ c_{39}=c_{31}r_7\ ,\ c_{40}
=c_{31}r_8$\hfil&\cr&&\hfil 12\hfil&&\hfil
$c_{41}=c_{33}r_2\ ,\ c_{42}=c_{33}r_6\ ,\ c_{43}=c_{34}r_5\ ,\ c_{44}=
c_{34}r_7\ ,\ c_{45}=c_{36}r_5$ ,\hfil&\cr&&&& \hfil$c_{46}=c_{36}r_7\ ,\ c_{47}=c_{37}r_7\ ,\
c_{48}=c_{37}r_8\ ,\ c_{49}=c_{39}r_6\ ,\
c_{50}=c_{39}r_8$\hfil&\cr&&\hfil 13\hfil&&\hfil
$c_{51}=c_{41}r_6\ ,\ c_{52}=c_{42}r_5\ ,\ c_{53}=c_{42}r_7\ ,\ c_{54}=c_{43}
r_7\ ,\ c_{55}=c_{43}r_8\ ,\ c_{56}=c_{45}r_4$ ,\hfil&\cr&&&&\hfil$ c_{57}=c_{45}r_7\ ,\
c_{58}=c_{45} r_8\ ,\ c_{59}=c_{47}r_6\ ,\ c_{60}=c_{47}r_8\ ,\ c_{61}=c_{49}
r_8$\hfil&\cr&&\hfil 14\hfil&&\hfil$c_{62}=c_{51}r_5\, ,\
c_{63}=c_{51}r_7\, ,\ c_{64}=c_{52}r_4\, ,\
c_{65}=c_{52}r_7\, ,\ c_{66}=c_{52}r_8$\ ,\hfil&\cr&&&&\hfil$
c_{67}=c_{54}r_6\ ,\ c_{68}=c_{54}r_8\ ,\     
c_{69}=c_{56}r_7\ ,\ c_{70}=c_{56}r_8\ ,\ c_{71}=c_{57}r_6$
,\hfil&\cr&&&& \hfil$c_{72}
=c_{57}r_8\ ,\ c_{73}=c_{59}r_8\ ,\ c_{74}=c_{61}r_5$\hfil&\cr&&\hfil 15\hfil&&
\hfil$c_{75}=c_{62}r_4\ ,\ c_{76}=c_{62}r_7\ ,\ c_{77}=c_{62}r_8\ ,\ c_{78}=c_{64}
r_7\ ,\ c_{79}=c_{64}r_8\ ,\ c_{80}=c_{65}r_6$ ,\hfil&\cr&&&&\hfil$
c_{81}=c_{65}r_8\ ,\ c_{82}
=c_{67}r_8\ ,\ c_{83}=c_{69}r_6\ ,\ c_{84}=c_{69}r_8\ ,\ c_{85}=c_{70}r_5\ ,\
c_{86}=c_{71}r_8$ ,\hfil&\cr&&&&\hfil$ c_{87}=c_{73}r_5\ ,\
c_{88}=c_{74}r_4$\hfil&\cr&&\hfil 16\hfil&&\hfil
$c_{89}=c_{75}r_3\ ,\ c_{90}=c_{75}r_7\ ,\ c_{91}=c_{75}r_8\ ,\ c_{92}=c_{76}
r_6\ ,\ c_{93}=c_{76}r_8\ ,\ c_{94}=c_{78}r_6$ ,\hfil&\cr &&&&\hfil$c_{95}=c_{78}r_8\ ,\
c_{96}=c_{79}r_5\ ,\ c_{97}=c_{80}r_8\ ,\ c_{98}=c_{82}r_5\ ,\ c_{99}=c_{83}
r_5\ ,\ c_{100}=c_{83}r_8$ ,\hfil&\cr&&&&\hfil$ c_{101}=c_{84}r_5\ ,\ c_{102}=c_{85}r_6\ ,\
c_{103}=c_{86}r_5\ ,\ c_{104}=c_{87}r_4\ ,\
c_{105}=c_{88}r_3$\hfil&\cr&&\hfil 17\hfil&&\hfil
$c_{106}=c_{89}r_7\, ,\ c_{107}=c_{89}r_8\, ,\ c_{108}=c_{90}r_6\, ,\ c_{109}=
c_{90}r_8\, ,\ c_{110}=c_{91}r_5\, ,\ c_{111}=c_{92}r_8\,$,\hfil&\cr &&&&\hfil$c_{112}=c_{94}r_5\,
,\ c_{113}=c_{94}r_8\, ,\ c_{114}=c_{95}r_5\, ,\ c_{115}=c_{96}r_6\, ,\ c_{116}=
c_{97}r_5\, ,\ c_{117}=c_{98}r_4 $\hfil&\cr&&\hfil 18\hfil&&\hfil
$c_{118}=c_{106}r_6\ ,\ c_{119}=c_{106}r_8\ ,\ c_{120}=c_{107}r_5\ ,\ c_{121}=
c_{108}r_5$
,\hfil&\cr&&&&\hfil$ c_{122}=c_{108}r_8\ ,\ c_{123}=c_{109}r_5\ ,\ c_{124}=c_{110}
r_6\ ,\ c_{125}=c_{111}r_5$\hfil&\cr &&\hfil 19\hfil&&\hfil$c_{126}=c_{118}r_5\ ,\ c_{127}=c_{118}r_8\
,\ c_{128}=c_{119}r_5\ ,\ c_{129}=c_{120}r_4\ ,\ c_{130}=c_{120}r_6$\hfil&\cr
&&\hfil 20\hfil &&\hfil$c_{131}=c_{126}r_4\ ,\ c_{132}=c_{128}r_4\ ,\ c_{133}=c_{129}r_6$\hfil&\cr
&&\hfil 21\hfil &&\hfil$c_{134}=c_{133} r_5$\hfil&\cr&& \hfil 22\hfil &&\hfil$c_{135}=c_{134} r_8$\hfil&\cr
\tablerule\noalign{\smallskip}
 }} $$
\centerline{The coset representatives $c_j$ are given recursively in terms of
the simple reflections $r_i$.}\end